\documentclass[11pt,letterpaper]{article}
\usepackage[utf8]{inputenc}
\usepackage[hyphens]{url}
\usepackage[round]{natbib}

\title{Evolutionary Genetic Engineering in the Indo-Pacific: Conservation, Humanitarian, and Social Issues}
\author{Floyd A.~Reed\\Department of Biology\\University of Hawai`i at M\=anoa\\ \texttt{floydr@hawaii.edu}}

\begin{document}
\maketitle

The Indo-Pacific region contains a unique mix of opportunities for the development and use of genetic-pest-management, gene-drive, and gene-drive-like technologies. Here I collectively refer to these technologies as Evolutionary Genetic Engineering (EGE). Indo-Pacific Islands have some of the world's highest rates of endemism and extinction—species and entire ecosystems are at risk. This threat to the natural world is coupled with the burden of human diseases, many of which are new and emerging or neglected tropical diseases. The same factors which have led to high rates of endemism also, in some ways, make this region an ideal testing ground for some types of EGE's. There is great potential for positive humanitarian, economic, and conservation applications of EGE's. However, these types of new technologies will be initially viewed from the perspective of the recent history of a loss of self determination, issues of social justice, and the testing of new technologies (e.g., biocontrol, agricultural, nuclear) in the Indo-Pacific—a region of the world that is still extensively colonized and controlled by Western Nations. Experience with successes and failures in related technologies suggests a path to move forward—a set of eight recommendations—to maximize the potential payoffs and minimize unintended negative effects of EGE's.

\section*{Introduction}
The Island Indo-Pacific is a large, important, unique, and unfortunately often overlooked region of the world. There is tremendous potential for the positive use of Evolutionary Genetic Engineering (EGE) in the region in both humanitarian and conservation applications. This potential stems from the regions geographic isolation, collection of infectious diseases, and species conservation urgencies. However, it would be a mistake to neglect the context of recent and ongoing political and social challenges in the region. Doing so is likely to generate a negative reaction that could inhibit the applications of promising emerging technologies. This context includes issues of colonialism, self determination, biocontrol, the testing of new technologies, and early experiences with genetically modified agricultural crops in the region. In this article I am focused on the Island Indo-Pacific Tropics, but also use examples from the broader region including India and Australia. I am also focusing on terrestrial applications of EGE’s. There are potential freshwater and marine applications, but this is less developed and goes beyond the scope of the current article.

In order to move forward in a way that does not sacrifice long term progress for short term convenience, we must accept that everyone has a role to play in shaping our technological future; this is not always easy to do when faced with confrontations and fundamental disagreements. To do this we must
\begin{enumerate}
 	\item enhance communications and avoid a reluctance to provide more detailed information about new technologies or to be dismissive of inquiries.
 	\item EGE applications should only be pursued if there is a genuine benefit to the local population (and if the people potentially affected generally agree that this is desirable rather than the decision being made externally), not in order to test new technologies in a “safe” manner or to avoid jurisdictional regulations.
 	\item The potential benefits and risks of EGE's, along with the degree of uncertainty surrounding both, need to be unambiguously communicated.
 	\item There needs to be a frank discussion of unintended side effects and the potential for misuse of the technology.
 	\item Humanitarian goals need to be administered and controlled by humanitarian organizations while conservation goals need to be administered and under the control of conservation organizations (applicable to both governmental and non-governmental organizations); this is especially true in an international setting.
 	\item Proactive research needs to be conducted and the data available to address common concerns about the possible ecological and health effects of EGE’s.
 	\item It takes broad perspectives, beyond what any single person is capable of, to identify potential promises and pitfalls of the development and implementations of EGE's.
 	\item Finally, a broad-based community discussion of, and direct involvement in, EGE development and applications should occur as early as possible. This will positively shape both the development and applications of the technology and help build a solid social foundation for future developments.
\end{enumerate}

\section*{Potential Evolutionary Genetic Engineering Applications in the Island Indo-Pacific.}
The Indo-Pacific spans half of the Earth’s circumference yet receives relatively less international focus. A revealing example is that this is the non-polar region that is most often divided on map projections of the world. This inattention is not simply due to a smaller population; the four most populous countries in the world (China, India, the USA, and Indonesia) have territory and active interests in the region.

The Island Indo-Pacific has some of the world’s highest rates of both species endemism (unique genetic diversity) and extinction \citep{vitousek1988, fleischer1998, myers2000, kier2009}. 
Adaptive radiations of species here have served as prime examples of evolutionary biology \citep[p.~380][]{darwin1845, dobzhansky1973}. 
Extinctions in these species-rich regions are proceeding at an alarming rate \citep{pimm1995, ganzhorn2001, fonseca2009, loehle2012, regnier2015} and this is predicted to be exacerbated by climate change \citep{benning2002, mora2013}. 
The region is in dire need of effective conservation strategies and potential Evolutionary Genetic Engineering (EGE) applications targeting introduced species and diseases have been proposed to establish effector genes refractory to introduced vectored disease (genetic modifications to block transmission of the disease) and genetic sterile insect techniques to suppress populations of invasive species \citep{clarke2002, wimmer2005, sinkins2006, altrock2010, esvelt2014, reeves2014, webber2015}.

The Island Indo-Pacific is also home to newly emerging and/or neglected tropical diseases that affect human health as well as economically important species. Vector borne human diseases in the region include chikungunya, dengue fever, Japanese encephalitis, lymphatic filariasis, malaria, plague and Rift Valley fever (in Madagascar), schistosomiasis, scrub typhus, West Nile fever, and zika. Additionally there are diverse agricultural crop pests and diseases that impact food production across the region. A major goal of EGE development is to address human disease, and there is also potential for agricultural applications \citep{alphey2002, sinkins2006, gould2008a, wimmer2013, esvelt2014, champer2016}.

When countries are listed by gross domestic product per capita it becomes apparent that the Indo-Pacific contains, in terms of national economic wealth, many of the poorest countries in the world. For example, the Comoros, Kiribati, Madagascar, Marshall Islands, Micronesia, Papua New Guinea, Solomon Islands, Tuvalu, and Vanuatu have an average per capita GDP of Intl.\$2,387, approximately 1/8th of the world average, Intl.\$18,872 \citep{IMF2016}. This limits the resources available that these governments can apply to humanitarian and conservation interventions and suggests an enhanced value of international collaboration.

In many ways the terrestrial isolation that has led to the Indo-Pacific's tremendous biological diversity also makes the region ideal for some EGE applications. Suppressing or modifying non-native invasive species is an obvious place to start. However, what may be a pest in one location may be a highly valued or important ecological species in another (e.g., nopal \textit{Opuntia} cacti are a highly valued component of Mexican cuisine and source of animal fodder while considered an invasive species pest in Australia---\textit{Cactoblastis cactorum} has been used successfully as bio-control in Australia but is now threatening native \textit{Opuntia} in the Americas \citep{zimmerman2004}. Proper application of Type 1 and 2a EGE’s (Appendix A: Types of Evolutionary Genetic Engineering) can leave a species genetically unmodified within its native range, even with low levels of migration between islands or islands and continents \citep{altrock2010, altrock2011, laruson2016}. The limited and discrete partitioning of land area of islands make 100\% local genetic transformation or eradication of a species possible without resorting to type 2b or 3 EGE’s (Appendix A) and allows the application to proceed in a stepwise fashion across multiple islands using limited resources.

\section*{Colonialism, self determination, and the testing of new technologies}
We have different perspectives depending on our experiences and social / cultural identities, and we are all-too-often not aware of how our individual perspective differs from others. In the middle of the abstract I used the following sentence, ``The same factors which have led to high rates of endemism also, in some ways, make this region an ideal testing ground for some types of EGE's.'' I chose the wording of this sentence carefully. What was your reaction? For many of the people reading this article the sentence seemed perfectly natural and flows into the ideas of the preceding and following sentences. However, for some readers the phrase ``testing ground'' is likely to stand out. Our reaction to this sentence is related to our perspective. For many who do not live in the island Indo-Pacific it is easy to see the region as something external to our daily lives and more disposable for testing and experimenting. In contrast, for some the Indo-Pacific represents home, family, work, and is also fundamentally connected to a cultural identity. I ask readers to construct your own sentence connecting a place that is highly valued to you personally (your hometown, where you live now, or a place of historical, religious, or cultural importance) with a ``testing ground'' of a new potentially powerful technology with its own set of concerns and unknowns.

The Indo-Pacific has a long and continuing history of a loss of self determination and sovereignty. 
The UN Special Committee on Decolonization lists American Samoa, French Polynesia, Guam, New Caledonia, Pitcairn, and Tokelau as Non-Self-Governing-Territories. 
The total number of ongoing sovereignty disputes encompasses many more islands and regions too extensive to list here. Colonization includes the establishment of extensive military bases and use of the islands for tests of nuclear, biological, and chemical warfare technologies---and these were not limited to a few isolated incidents, for example hundreds of nuclear weapons tests were conducted by France in Mururoa and Fangataufa Atolls, by the United Kingdom in South Australia, Montebello, and Kiritimati Islands, by the United States in Pikinni (Bikini), \=Anewetak (Enewetak), Johnston (Kalama) Atolls, and Kiritimati. 
This history of military testing, non-military testing of new technologies (e.g., disastrous attempts at classical biological control by introducing new species, e.g., \citealt{howarth1983,clarke1984,henneman2001,messing2006,hays2007,parry2009}), and colonization in the region can severely inhibit international biological research and potential applications including EGE’s.

There is a case study that deserves special mention within the context of EGE’s in the Indo-Pacific, especially in the context of international programs and applications of mosquito genetic engineering. 
From 1969–1975 the World Health Organization (WHO) collaborated with the US Public Health Service (PHS) and the Indian Council of Medical Research (ICMR) to establish a Genetic Control of Mosquitoes Research Unit (GCMRU) in India; this was financially supported by the Government of India, US PL-480 funds, and the CDC \citep{AEND1975a}. 
The GCMRU was studying and implementing mosquito control technologies including the release of sterilized individuals. What appears to have started with concerns about a carcinogen (thiotepa) being added to well water in the village of Pochanpur without public or government consultation got caught up in politics \citep{AEND1975b,hanlon1975}, with widespread accusations in the media and later by the Government of India, and grew into a political disaster with suspicions that the US military was using India to test methods of biological warfare using mosquitoes \citep{AEND1974, sehgal1974,anonymous1975, AEND1975b, hanlon1975, anonymous1976,powell2002}. 
The addition of thiotepa to village water has been denied by WHO \citep{tomiche1975}, but publications preceding the accusations suggest this may have happened \citep[pp. 85-87][]{pal1974}---and therein lies one problem. 
There was a lack of clear unambiguous communication from the beginning. Furthermore, PHS did have military connections and shared materials and information with the US military \citep{langer1967,treaster1975}. 
The US military did conduct chemical and biological tests in the Indo-Pacific; this included the release of mosquitoes off the coast of Baker Island (``Magic Sword'' 1965), the release of \textit{Bacillus globigii} in O`ahu (``Big Tom'' 1965), shelling sarin nerve agent in Wai\=akea Forest Reserve, Hawai`i (``Red Oak'' 1967), and the dispersal of \textit{Staphylococcus aureus} enterotoxin type B over \=Anewetak (Enewetak) Atoll (``DTS Test 68-50'' 1968). 
However, in all likelihood there was no military or biological warfare connections with the GCMRU \citep{WHO1976,powell2002}. 
Covert transfer of US funds to keep GCMRU going was briefly discussed with WHO \citep{AEND1975c, SSWDC1975} but this was considered too risky and the US suspended funding the project. 
Despite denials by WHO \citep{tomiche1975}, the GCMRU, which was planned to extend at least until 1978, was forced to shut down prematurely in 1975 \citep{AEND1975d} and the project was deemed a failure \citep{curtis2007}.

What can be learned from this?
\begin{enumerate}
 	\item There was a clear lack of communication resulting from a reluctance of either the WHO or the US to engage the media and comment on the allegations \citep{AEND1974, AEND1975e, anonymous1975, tomiche1975}. This was unfortunate as it, perhaps rationally, fuelled suspicions. The public perception of public perception may differ from public perception---the individual perception of public opinion is influenced by a range of factors and may not be an accurate reflection of commonly held attitudes \citep[e.g.,][]{mutz1989}. The idea that providing more information would undermine support conflicts with recent results that show the more informed people are of the release of genetically modified mosquitoes the more supportive they become; however, a great deal of public engagement has to be accomplished, especially for women, minorities, and people with lower education levels and lower household incomes \citep{ernst2015,kolopack2015}. 
 	\item There was a perception that these experiments would not have been permitted in Western countries and that India was being used as a testing ground \citep{anonymous1975,raghavan1975}. Knowledge that a technology has been effective in other countries is one factor associated with strong public support \citep{ernst2015}. It is unfortunate that prior programs in the US, Myanmar/Burma, Tanzania, Western Africa, and France were not communicated to the Indian press \citep{laven1972,WHO1976,curtis2007}. The perception that an international project is being conducted to avoid home country regulation should certainly be (truthfully) avoided. 
 	\item The potential benefits of the project to the people of India was unclear \citep{AEND1974,anonymous1975}. This is perhaps most tragic of all. India suffers from mosquito vectored dengue, malaria, Japanese encephalitis, chikungunya, and lymphatic filariasis \citep{sharma2015}. While a balance must be struck to not over-promise results that may not be realized, the goals and potential benefits of EGE applications must also be clearly advertised. 
 	\item There was a lack of an \textit{a priori} open and frank discussion about possible misuse of the technology \citep{hanlon1975}. While any technology can be potentially misused by individuals or organizations, a nation's government, and especially its military, has non-humanitarian and non-conservation priorities that can potentially conflict with the goals of humanitarian and conservation projects. Regardless of the existence of an actual conflict, the perception of possible conflict does exist, which can undermine credibility \citep{serafino2008,charney2013}. Fortunately today this is widely recognized and the 1978 UN ENMOD treaty (\url{http://www.un-documents.net/enmod.htm}) may prevent, depending on interpretation, military involvement in EGE technologies except perhaps for some limited applications of type 0 and 1 systems (Appendix A). The ENMOD treaty states that ``Each State Party to this Convention undertakes not to engage in military \ldots environmental modification techniques having widespread, long-lasting or severe effects \ldots the term `environmental modification techniques' refers to any technique for changing---through the deliberate manipulation of natural processes---the dynamics, composition or structure of the Earth, including its biota \ldots'' However, there is still a need for an open discussion about potential malicious uses, and military involvement with EGE projects should be avoided in order to encourage international trust and cooperation. 
\end{enumerate}
In response, Dr.~B.~D.~Nagchaudhuri's, physicist and scientific adviser to the Indian Ministry of Defence, recommendations were ``(A) that research proposals and projects are available to the public; and (B) that pertinent records contain clear statements as to why the objective is important, what is the [Government of India's] interest, and what is the [United States Government's] interest'' \citep{AEND1975f} and ``ministry officials must be alerted to any sensitive problems by the technical experts involved''; also, that ``each collaborative project should also be approved at the ministerial or secretary level of the ministry under which the project would fall i.e. health projects - Ministry of Health, Agricultural Projects - Ministry of Agriculture, \ldots This should also hold true whether on the Indian side or the US side'' \citep{AEND1975g}. Dr.~Hanlon recommends ``At the very least, there should be an open discussion of the [biological warfare] potential of such projects before they begin, so that countries can make informed choices'' \citep[p.~103][]{hanlon1975}.

There is a value to compartmentalizing different aspects of a government’s actions. It seems almost self evident that funds for research are best spent by research agencies, funds for health are best spent in agencies focused on health, funds for conservation are best spent by agencies trained in and focused on conversation. Even if there were sufficient funding and resources we would not want the EPA (Environmental Protection Agency) or DOH (Department of Health) carrying out military actions; the converse is also true. We don’t want to rely on our military to carry out conservation, human health, or humanitarian actions when there are other agencies, without conflicting priorities, that can and should be doing this \citep{serafino2008,charney2013}. The author has discussed EGE's and the ENMOD treaty in person with current and former members of DARPA, a research branch of the military with an interest in EGE's, and has been told that the military has to carry out high risk (in the sense of new and experimental) research because NSF (National Science Foundation) and NIH (National Institutes of Health) cannot. I completely disagree. Research agencies can and should also be funding higher risk, higher pay-off research instead of abdicating this role to the military---and to avoid the kinds of conflicts suggested in the WHO experience in India. This is not done in the US because of historical inertia and objectively unbalanced federal budget allocations (a Department of Defense, DOD, estimated research budget of \$66 billion versus \$29 billion for NIH and only \$6 billion for NSF in FY2015, \citealt{hourihan2016}). Reallocating civilian research funds to civilian agencies would also free up the military to focus on military actions and capabilities.

\section*{Recent experiences with GMO’s in the Indo-Pacific}
EGE’s are likely to be initially framed in terms of the GMO (Genetically Modified Organism) crop debate \citep{knols2007}.\footnote{Although, classical sterile insect technique involving radiation or chemicals and Wolbachia based techniques stand as exceptions. These are genetic approaches in the sense that the organisms chromosomes are affected (Callaini et al., 1997; Robinson, 2005), but they are not considered genetic modifications.} Within Hawai`i, Rainbow Papaya and GMO Taro serve as contrasting examples of the interaction between social acceptance, development, and deployment of new technologies. \textit{Carica papaya} was not grown in Hawai`i until after European contact in 1778. The papaya industry in Hawai`i was devastated in the 1990’s by the ringspot virus. A genetically engineered ``rainbow'' papaya resistant to ringspot infection was developed at Cornell University by Dr.~D.~Gonsalves \citep{ferreira2002} who was originally from Hawai`i. While GMO papaya is not without controversy \citep[e.g.,][]{harmon2014,hofschneider2016} it is credited with rescuing the industry and is \textit{de facto} widely adopted in Hawai`i today \citep[e.g.,][]{kallis2013}.

\textit{Colocasia esculenta} (Taro or Kalo in Hawaiian) was brought to Hawai`i by the ancient Polynesians. A wide range of Kalo varieties have had a central role in traditional Hawaiian culture as a staple food crop and continues to be economically important \citep{whitney1939,fleming1994}. Furthermore, Kalo is literally the brother of humans (H\=aloa) in the Hawaiian creation tradition and words for family and relationships also refer to parts of the plant \citep{kahumoku1980}. Taro leaf blight (\textit{Phytophthora colocasiae}) was introduced to Hawai`i in the 1900’s and has significantly impacted Kalo \citep{nelson2011}. Work at the University of Hawai`i was begun to to breed resistant varieties which resulted in patents in 2002. Separately a Chinese variety of Taro was genetically modified from 2001 to 2006 with a gene from wheat to be resistant to leaf blight. This resulted in widespread public outrage and large protest rallies in 2006 that resulted in the university relinquishing its patents and issuing an indefinite moratorium on the genetic engineering of Hawaiian Kalo \citep{ritte2006,CTAHR2009}.

With these cases in mind consider a potential EGE project. Culex mosquitoes were introduced to Hawai`i in the mid 1800’s. They vector \textit{Plasmodium relictum} which is responsible for avian malaria. Many Hawaiian forest bird species, important in traditional Hawaiian culture (e.g., `ahu `ula, mahiole, and in Hawaiian religion), have no immunity or tolerance to \textit{P.~relictum} and have become extinct, with many currently threatened, as a result \citep{warner1968}. These two previous contrasting examples suggest that genetically modifying non-native mosquitoes to reduce the frequency of avian malaria is much more socially acceptable than the reverse: genetically modifying native Hawaiian birds to be resistant to infection by Plasmodium (although it would be worth conducting the relevant public surveys to determine this). Also, doing the research locally in Hawai`i is not necessarily an advantage in terms of securing broad local public support, buy-in, and acceptance (however, it is an advantage in terms of engaging the public).\footnote{Another fascinating dimension is the degree of public awareness and identity with conservation goals and issues. This varies tremendously across the Indo-Pacific with various emphasis on terrestrial and marine issues and could be the subject of an article and research project in its own right.} These are aspects that might not initially be appreciated by scientists designing EGE technologies.

On a broader scale across the Indo-Pacific, consider the cases of golden rice and Bt-cotton. Rice (\textit{Oryza sativa}) is a staple crop for a large segment of the population across the Indo-Pacific. A major nutritional shortcoming of rice is the lack of beta-carotene that can be metabolized into vitamin A, which in many of these populations is \textit{de facto} not simply rectified by supplementing with additional food sources. This unfortunate situation leads to blindness and the deaths of over half a million people a year. To address this, rice has been engineered since 2000 with DNA sequences from other plants to produce bio-available beta-carotene \citep{ye2000,paine2005,tang2009}. This ``golden rice'' has also been the target of a great deal of controversy, protest, and misinformation \citep[e.g.,][]{dobson2000,potrykus2001,enserink2008,lynas2013b}. Much of this protest originates in the Western world where ironically we have a wide range of nutritional supplements added to our food including vitamin D in milk, calcium in orange juice, niacin and folic acid in bread, iodine in salt, and fluoride in drinking water. One question to ask ourselves is, why is it so easy to add all of these supplements to our food supply, not to mention widespread adoption of genetically modified corn, soybeans, cotton, potatoes, sugar beets, \textit{etc}., in parts of the West, when providing vitamin A in the form of Golden Rice for much of the world’s population is still not approved and remains in a testing phase well over a decade later?

Bt-cotton, which has received less attention in the media, provides a contrasting case to golden rice where a GM crop has been embraced in the Indo-Pacific and this has been in large part driven by local buy-in. Bt-cotton is engineered to produce a naturally occurring insecticide from a bacteria (\textit{Bacillus thuringiensis}). The intention is to kill larvae of the cotton bollworm (\textit{Helicoverpa armigera}). A seed company in India led by Dr.~D.~B.~Desai began selling ``Navbharat 151'' seed in 1998 with the claim that the plants did not have to be sprayed with pesticides for bollworm. This proved to be the case during a large bollworm outbreak in Gujarat in 2001, which raised questions. It was found that Navbharat 151 plants had a genetic modification created by Monsanto. The Indian government filed criminal charges against Dr.~Desai, ordered the seed destroyed, and 4,000 hectares of planted fields burned. Thousands of farmers rallied to support Dr.~Desai and block burning the fields; the Gujarat government refused to carry out the order; the recall was cancelled, and some farmers saved their own seed for replanting. The opposite of concerns about using India as a testing ground as discussed in the WHO mosquito project of the 1970's (point 2 above) were expressed: `How can something made in the United States, many of them wonder aloud, be unsafe in India? ``I think they grow it in China and other countries,'' says Kalidas Patel, who grew Navbharat cotton in Gujarat' \citep{mcgray2002}. Later Monsanto was granted a license to market Bt-cotton in India and in all likelihood the prior experience with Navbharat 151 promoted public buy-in \citep{menon2001,mcgray2002}. In recent years Bt-cotton is widely adopted, approximately 90\% of the cotton grown in India, and a black market for Bt-cotton seeds also appears to be thriving \citep[e.g.,][]{kathage2012,nemana2012}. 
However, this is in no way a simple matter and debates regarding Bt-cotton, Monsanto, and regulation continue \citep[e.g.,][]{anonymous2016,basheer2016}. 
Regardless, the support among Indian farmers for Bt-cotton stands in stark contrast to the protests over golden rice being planted in test beds in the Philippines \citep{lynas2013a}. The cause of the difference between these experiences is hard to isolate and a large number of idiosyncratic effects likely contribute including the pivotal actions of a few or a single individual. However, the effects of local buy-in, combined with local access to technologies, and first hand experience with these technologies, should not be ignored.

Finally, concerns about ecological effects of EGE’s are associated with strong opposition to the technology \citep{ernst2015}. There are also questions of possible, but unlikely, bioaccumulation of toxic proteins and allergenicity \citep{curtis2007,reeves2012}. In addition to the four guidelines in the previous section, despite limited time and funding, we should conduct the work to have the data on hand to address these questions to the public \citep{curtis2007}.

\section*{Everyone has a role to play}
We live in a world that is often overly self-polarizing. 
I am a geneticist; I entered this field because of personal interest, excitement, and challenges of the promise and potential of genetics. 
Unintentionally, this has become a part of my identity. 
When I was first exposed to protests over genetic technology it was all too easy to feel that it was also a personal attack. This is nested within the context of broader anti-scientific popular views related to climate change, evolution, renewable energy, vaccinations, \textit{etc}. The natural reaction is to reflexively move in the opposite direction and argue that genetic technologies are safe, protesters don't understand the issues, \textit{etc}. and be overly dismissive; a position that I may not have had initially. 
The difficult but essential step for growth is to try to find a middle ground and synthesize a path forward \citep[see also][]{NPR2013}. Right or wrong, no single perspective can do this on its own and, because of our perspectives, we are often blind to potential issues apparent to other people. It is easier to see a potential risk if you are looking for a risk instead of working toward developing a desired application of a new technology. For example, the potential of allergic reactions to genetic modifications are real and not to be dismissed \citep[e.g.,][]{nordlee1996}, and many crops have a strong cultural significance that many people may not be aware of such as Kalo in Hawai`i, discussed above, or maize in Chiapas \citep{bellon1994,perales2005,brush2007}. As geneticists we are in a unique position to be able to critically assess potential benefits and risks, once we perceive them, of genetic technology from a scientific perspective. It is our responsibility to embrace and communicate this rather than contributing to destructive polarization. However, it is not our job to be overly encompassing and give equal weight to all objections; we also must be willing to learn from past experiences (such as the disastrous effects of the perception of possible military involvement in the WHO program in India) and to rationally disagree when we reason this to be the case. For example, despite claims to the contrary \citep{seralini2012}, there is no scientific evidence that herbicide resistant maize is carcinogenic. There is a great deal of misinformation and misconceptions surrounding who would or would not benefit, and to what degree, from golden rice \citep{harmon2013}. Attitudes regarding GMO's are divisive, some are not based on factual evidence and can be labeled as irrational although this quickly gets complex \citep{stone2010,lynas2013a,blancke2015,hicks2015}; regardless, the GMO debate will continue to prove a rich subject for the analysis of the dynamics of politics, the media, framing effects, confirmation bias, social identity, information cascades, \textit{etc}., for many years to come.

An area that can benefit from improvement is to incorporate this synthesis earlier into the research and development process. If individuals with different perspectives were able to directly participate in the design of a new technology, they could shape the direction in which it develops towards an outcome that might be more desirable and socially acceptable. (Recall the effect of personal experience with Bt-cotton and its adoption in India.) Often the way development of a new technology works is in incremental steps of design, troubleshooting, and research funding, to consultation and approval from regulatory agencies, to building the logistics of application and deployment. Public consultation and asking for acceptance occurs only at the end of the day, when many steps have been cast and it is more difficult and time consuming to make fundamental revisions. One possibility is to include grant support for individuals from the social sciences to be ``embedded'' in a biological laboratory in order to fully participate in a laboratory’s research and conduct their own research about social attitudes, context, communication, perceptions, \textit{etc}., both in its own right and as a bidirectional conduit to facilitate communication, public guidance, and knowledge transfer in the development of EGE technologies (see \citealt{kolopack2015} for a highly effective example of community engagement albeit not exactly in the same form that I am proposing here). The local community can directly participate in the development of a new technology, possibly facilitating progress in a direction that is unanticipated by the researchers, funding, and regulatory agencies, but one that results in a greater positive potential being realized at the end of the day.

\section*{Conclusion}
This Indo-Pacific is geographically isolated, under a burden of infectious diseases, and is in dire need of protection of its natural world. This creates an opportunity for positive, highly valued, effective applications of EGE's. However, it would be a mistake to ignore the history and social realities thought the region. To reiterate the eight points from the introduction that have been expanded upon through this article:
\begin{enumerate}
 	\item There is a need to enhance and engage communications in all directions.
 	\item EGE applications should only be pursued if there is a genuine benefit to, and buy-in from, the local population. 
 	\item The potential benefits and risks of EGE's need to be unambiguously communicated. 
 	\item There needs to be a clear unambiguous discussion of unintended side effects and potential misuses of the technology. 
 	\item Humanitarian goals need to be administered and controlled by humanitarian organizations and conservation goals need to be administered and under the control of conservation organizations. 
 	\item Proactive research needs to be conducted and the data available to address common concerns. 
 	\item It takes broad perspectives to broadly identify potential promises and pitfalls of EGE's. 
 	\item An early broad-based community discussion of, and involvement in, EGE development and applications should occur. 
\end{enumerate}
Finally, no matter how ``new'' a technology or situation seems, there is still much to be learned from history.

\section*{Ancknowledgements}
I thank \'Aki L\'aruson, Vanessa Reed, and three anonymous reviewers for comments on the manuscript. Related work in the Reed lab has been recently supported by the Hawai`i Community Foundation, the State of Hawai`i Department of Land and Natural Resources, and by the Reed family. 

\section*{Appendix A: Types of Evolutionary Genetic Engineering}
An important concept that cannot be over-emphasized is the diverse types of EGE's and their predicted effects. At the risk of oversimplification, here are four main types with an important boundary between them.\footnote{Another important type of classification are the types of likely effects and dynamics that occur when the drive systems are disrupted by mutation, recombination, and selection and how EGE's are likely to be converted into different types, but this goes beyond the scope of this article.}
\begin{description}
	\item [Type 0]
 	Generic genetic modifications not designed to change in frequency over time using evolutionary principles. In general these are expected to either drift neutrally (if there is little to no effect) or be removed by natural selection. For example fluorescent proteins are often used to mark and keep track of genetic inserts; however, these proteins can have toxic effects \citep[e.g.,][]{liu1999, devgan2004, shaner2004, shaner2005}. This tends to reduce an organism’s fitness and these modifications are not expected to persist in the wild over many generations.
 	\item [Type 1]
 	Deleterious EGE's that are designed to be transient and removed from the population. Examples of type 1 include the “killer-rescue” system \citep{gould2008b}, genetic sterile insect technique \citep{horn2003}, and Wolbachia in cytoplasmic incompatibility population suppression applications \citep{laven1967, knipling1968}. These may persist in the wild for a shorter period of time than type 0 EGE’s.
 	\item [Type 2]
 	Threshold EGE’s that cannot increase in frequency when very rare but can increase in frequency and persist indefinitely once a critical frequency point is passed.
\begin{description}
 	\item [Type 2a]
 	Thresholds that are above a frequency of one half. These include chromosomal rearrangements \citep{foster1972}, haploinsufficient induced underdominance \citep{reeves2014} and possibly some forms of maternal-effect underdominance \cite{akbari2013}.
 	\item [Type 2b]
 	Thresholds that are below a frequency of one half. This includes Wolbachia \citep{hoffmann2011}, some forms of maternal-effect underdominance \cite{akbari2013}, and some theoretical systems \citep{davis2001}.
\end{description}
 	\item [Type 3]
 	Unconditionally driving EGE’s that can invade a population from arbitrarily low frequencies. These include Medea \citep{chen2007}, homing endonucleases \citep{windbichler2011}, transposable elements \citep{carareto1997}, meiotic drive \citep{cha2006} and some types of CRISPR systems \citep{gantz2015, dicarlo2015, hammond2015}.
\end{description}
In one perspective, the most important distinction is the boundary between 2a and 2b. This predicts what will happen without human intervention (without additional releases of modified or unmodified individuals) among multiple populations within a species due to the forces of migration and selection \citep{barton2011}. Type 1-2a will tend to reduce in range and disappear (although this may take many generations) while type 2b and 3 will tend to spread and become more established (and this may occur in a small number of generations for type 3) with the concern that once widespread enough this may be irreversible. While type 2a systems might be considered ``gene drive'' in a broad sense the term is probably more accurate to describe type 2b and especially type 3 systems (gene drive in the strong sense). The boundary between 2a and 2b represents a balance between ease of transformation of a population and reversibility back to a transformation free state---a balance between safety and efficiency.

Some natural EGE systems in the type 3 category have been shown to be capable of moving across subspecies and species boundaries, rapidly spreading worldwide, and lowering the average fitness of a species \citep[e.g.,][]{eanes1988, morita1992, hill2016}. The concern of this possibly happening due to artificial genetic engineering is not a new one \citep{gould2006}). For example, fully functioning transposable elements have been introduced into various new species in the lab \citep[e.g.,][]{brennan1984, daniels1989}, sometimes with little to no discussion of containment and possible escape. Fortunately there are methods of building in safeguards to minimize the chance of unintended spread in the wild \citep{dafaalla2006, gokhale2014}.

Additionally, it is important to keep in mind the (sometimes unexpected) effects of mutations and selection that can change the dynamics of EGE’s. For example, Y chromosome meiotic drive can be quickly suppressed by sex chromosome aneuploidy \citep{lyttle1981}. Arthropod species have been observed to rapidly evolve to suppress some effects of Wolbachia \citep{charlat2007}. Type 0 EGE’s may drift at some frequency in a population by unintended contamination \citep[e.g.,][]{gonsalves2012, xiao2016}; one concern that goes beyond this is that genetically engineered disease resistance may be adaptive, if infection by the disease has a large enough fitness cost, and the type 0 EGE may deterministically increase in frequency in the wild, essentially becoming a type 3 EGE (although to date there are not clear examples of this, e.g., \citealt{fuchs2004}). Some of these unexpected effects can be detected in laboratory experiments and incorporated into the design and predictions of the EGE.

It is already a challenge to filter out misinformation and misconceptions regarding genetic modifications. The author realizes that this adds another challenge; however, the fact is there are various types of EGE's with a range of predicted effects regarding how well they can be established in the wild and how reversible they are. It is appropriate, if possible, for these dynamics to be considered and to interact with regulatory approval and public acceptance \citep{harmon2014}.



\end{document}